\documentclass[pre,amssymb,aps,floatfix,nofootinbib]{revtex4}
\usepackage{graphicx,epstopdf}
\usepackage{dcolumn}% Align table columns on decimal point
\usepackage[sort&compress]{natbib}
\usepackage{bm}
\usepackage{subfigure}

\usepackage[dvips]{epsfig}

\usepackage{xcolor} 
\usepackage[normalem]{ulem}

\usepackage{bm}
\usepackage{color}
\usepackage[dvips]{epsfig}
\usepackage{amsmath} 
\usepackage{amssymb}

\newcommand{\comment}[1]{}

\newcommand{\BEA}{\begin{eqnarray}}
\newcommand{\EEA}{\end{eqnarray}}

\newcommand{\bq}{\begin{equation}}
\newcommand{\eq}{\end{equation}}
\newcommand{\be}{\begin{eqnarray}}
\newcommand{\ee}{\end{eqnarray}}
\newcommand{\ba}{\begin{align}}
\newcommand{\ea}{\end{align}}

\renewcommand{\d}{{\rm d}}

\renewcommand{\a}{\bm{a}}
\renewcommand{\b}{\bm{b}}

\renewcommand{\aa}{\hat{\bm{a}}}
\newcommand{\bb}{\hat{\bm{b}}}

\newcommand{\x}{\bm{x}}
\newcommand{\y}{\bm{y}}
\newcommand{\z}{\bm{z}}

\newcommand{\T}{\mathfrak{T}}
\newcommand{\na}{\bm{\nabla}}

\begin{document}

\title{Nonlocal, nonlinear electrostatics: like-charge attraction and asymmetric solvation}

\author{V. Stepanyan}
\author{Y. Sh. Mamasakhlisov}
\affiliation{Yerevan State University, Yerevan, Armenia}
\author{A. E. Allahverdyan}
\affiliation{Alikhanyan National Laboratory (Yerevan Physics Institute), Yerevan, Armenia}

\begin{abstract} 
We study an equilibrium system, where linear
molecules with a rigid charge distribution interact via the mean-field
generated by the Coulomb coupling. We focus on the nonlinear regime,
which is valid close to external charges immersed in the system.
The finite molecular size in this regime leads to a nonlocal response at short distances. First, we show that the mean-field potential is non-monotonic, i.e., 
like charges can attract each other. Second, we show that the polarization induced by an external charge (and hence its solvation structure) depends on the sign of this charge, violating the effective C-symmetry of the macroscopic electrodynamics.
We identify several scenarios of this effect, characterize it with the free energy and entropy differences, and compare it with analogous effects in asymmetric electrolytes described by nonlinear Poisson-Boltzmann equations.
\end{abstract}

\maketitle

\section{Introduction}

\subsection{Electrodynamics: macro {\it versus} micro}

Macroscopic electrodynamics provided the basis for microscopic electrodynamics \cite{landau8,cho,vlasov}. The opposite is now true: macro is derived from micro under several assumptions \footnote{We do not discuss assumptions related to characteristic times, since we plan to focus on electrostatics.} \cite{landau8,cho,vlasov}: {\it (i)} length-scale separation: the theory operates on an intermediate scale $\ell$ with $\ell_{\rm mol}\ll \ell\ll\ell_{\rm macro}$, where 
$\ell_{\rm mol}$ and $\ell_{\rm macro}$ are (resp.) 
molecular and macroscopic scales. {\it (ii)} The response 
of the matter to external fields is space-local: the external fields changing on scale $\ell$ induce a response on the same scale $\ell$. 
As a result of {\it (i)} and {\it (ii)}, we can neglect the inhomogeneity at scales smaller than $\ell$ and apply spatial averaging \cite{landau8}. Another consequence of {\it (i)} and {\it (ii)} is that molecules can be modeled as point-like dipoles. {\it (iii)} The linearized response of the matter to fields. {\it (iv)} If the matter is in equilibrium, and the external fields are time-independent, the problem moves from electrodynamics to electrostatics. 

Altogether, {\it (i)--(iv)} explain why features of the matter enter into macroscopic electrostatics mostly via the electric permittivity \cite{landau8}. Microscopic
theories attempt to constrain the permittivity via models, and go beyond the assumptions {\it (i)--(iv)}
\cite{landau8,cho,vlasov,dpb1,dpb2,dpb3,andelman-review,eps-stability,buyukdagli,kivelson,ralf,vanik}.
For example, {\it (i)} is relaxed for X-rays physics, whose wave-length
is comparable with inter-molecular distances (i.e., $\ell\sim\ell_{\rm
mol}$) \cite{landau8}. {\it (ii)} is relaxed in recent theories of non-point-like dipoles \cite{landau8,dpb1,dpb2,dpb3,buyukdagli}. {\it (iii)} is
relaxed within nonlinear electrostatics
\cite{landau8,vlasov,bavaud,vanik,boyd}. 

Our purpose is to study the nonlinear electrostatics of a molecular system
(aprotic liquid or a dense gas), where assumptions {\it (i)--(iii)} do not apply \footnote{Assumptions {\it (i), (ii), (iv)} (but not {\it (iii)}) are revised in the companion paper Ref.~\cite{stepanyan2026}. }. In particular, {\it (i)} does not apply, since we are interested in distances
comparable to intermolecular ones: $\ell\sim\ell_{\rm mol}$. Such
distances are relevant for electrostatic effects in biomolecular
reactions \cite{electrostatic_proteins}.  For the linearized, mean-field
theory of systems without {\it (i)} and {\it (ii)}, see
\cite{buyukdagli,ralf}. Ref.~\cite{dpb3} discusses susceptibilities of
weakly-nonlinear theories. We focus on uncovering new effects, which do
not exist in local or linearized theories. 

\subsection{Our assumptions}

Instead of {\it (i)--(iii)} we assume:

{\it (1)} Each molecule is described by point charges connected with rigid bonds (covalent bonds) \cite{hirshfeld}; i.e., each molecule is a classical rigid body from the viewpoint of charge distribution. This assumption simplifies the theory, and it agrees with how the molecular charge
distribution is described in the chemistry of dipolar compounds, and, in particular, zwitterions \cite{hirshfeld,hirshfeld2}. \footnote{A somewhat more general assumption (which is implementable, but will not be pursued in this paper) is that the charge distribution is delocalized, but fixed.} Note that the rigid-charge distribution is a simple model assumption. Its main limitation is that the induced dipole moment is neglected. For molecular systems, there are much more involved (also quantum) models \cite{water_models}. Due to technical demands of solving nonlinear integro-differential equations (with retarded and advanced arguments), we use in this paper even more restrictive assumptions on linear rigid molecules, where the effective molecular charges are located along a single line. 

{\it (2)} The matter distribution can be described within the mean-field (but not linearized!) approach. This assumption seems reasonable for dipolar gases and aprotic dipolar liquids. Note that {\it (2)} (and to a lesser extent {\it (1)}) is limited at sufficiently large densities. We understand that the validity limits of {\it (1)} and {\it (2)} overlap with nonlinearity and nonlocality; see the last section for a more targeted discussion.

\subsection{Our results}

Our main results are as follows. First, we show that there are
situations, where like charges (e.g., a positive external charge and a
positive test charge) attract to each other.  The origin of this effect
is the nonlocal polarization distribution around the external charge.

Second, we show that the effective charge-inversion symmetry (or
${\rm C}_{\rm eff}$-symmetry) can be effectively violated due to nonlinear and nonlocal effects via several different scenarios. The ${\rm C}_{\rm eff}$-symmetry means that the polarization (or solvation) structure induced by an external charge immersed in an electroneutral molecular system, does not depend on the sign of this charge. Equivalently, it means that the internal electric field changes only in sign in response to the sign change of the external charge. The effect of asymmetric solvation is well known \cite{asymmetric_solvation_0,asymmetric_solvation_1,asymmetric_solvation_2,asymmetric_solvation_3,asymmetric_solvation_4}. However, to our knowledge, it has not yet been related to nonlinear, nonlocal electrostatics. Instead, it was studied via molecular dynamics and phenomenological approaches, mostly for water molecules (i.e., for hydration) \cite{asymmetric_solvation_0,asymmetric_solvation_1,asymmetric_solvation_2,asymmetric_solvation_3,asymmetric_solvation_4}. 

As confirmed below, this symmetry holds for the macroscopic electrodynamics either in the linearized regime or in the nonlinear case under space-locality, or for dimer molecules under arbitrary nonlocality and nonlinearity. We identified two mechanisms of the breakdown of ${\rm C}_{\rm eff}$-symmetry in the dense gas (or liquid) of linear, asymmetric molecules. The first mechanism is exemplified by nitrosyl fluoride ($F-N=O$). This molecule has a permanent dipole moment with different densities of positive and negative charges, i.e., its charge distribution is asymmetric. The second mechanism is shown by carbon dioxide ($O=C=O$), a symmetric molecule with symmetric charge distribution and zero permanent dipole moment. 

Violations of ${\rm C}_{\rm eff}$-symmetry are also present for asymmetric electrolytes, as described by the nonlinear Poisson-Boltzmann mean-field equation. 
Here this symmetry is violated without requiring nonlocality. An example of such electrolytes is ${\rm Ca}^{+2}+2{\rm Cl}^{-}$, which is electroneutral, but contains ions with different charges (both in sign and absolute value). Despite the interest in the nonlinear Poisson-Boltzmann equation (see \cite{tellez_review,gray_review,xing} for reviews), the phenomenon of ${\rm C}_{\rm eff}$-symmetry violation did not (to our knowledge) get adequate attention in the literature. We show that the scenarios of the ${\rm C}_{\rm eff}$-symmetry violations in the nonlinear Poisson-Boltzmann system are different depending on the dimension (1 versus 3), i.e., depending on whether a point-like external charge or a charged plate are screened. 

\subsection{Structure of the paper}

The next section describes the mean-field
approach. Section \ref{section_dimers} works out this approach for dimer
molecules, and shows attraction of like charges. Limitations of the mean-field approach for dipolar systems are discussed in Appendix \ref{validity}. Section
\ref{section_generalized_dimers} discusses the ${\rm C}_{\rm eff}$-symmetry
and its violation. This effect is absent when one of its main
ingredients, nonlocality or nonlinearity, is absent. The ${\rm C}_{\rm eff}$-symmetry violation is studied with two representative examples of linear molecules: asymmetric dimers and symmetric molecules with zero dipole moment. We also provide a thermodynamic characterization of the ${\rm C}_{\rm eff}$-symmetry violation via free energy and entropy; see section \ref{free}. To gain a deeper understanding of the ${\rm C}_{\rm eff}$-symmetry and its limitations, we studied the mechanisms of its violation for asymmetric electrolytes within the nonlinear Poisson-Boltzmann equation; see section \ref{poisson}. 
We summarize our results in section \ref{summary}. 
The last section discusses numerous open problems, and (in this context) also limitations of the present work. Appendix \ref{free_ap} calculates the mean energy and free energy of dielectrics from statistical physics of mean-field systems. It also emphasizes the meaning of the variational free energy as the isothermal work. Appendix \ref{apa} presents the perturbative solution to the three-dimensional nonlinear Poisson-Boltzmann equation. Appendix \ref{apb} discusses the exact solution of the planar Poisson-Boltzmann equation for asymmetric electrolytes.

\section{The mean-field approach}

The mean-field treatment of $N$-body classical system is an old and
efficient method \cite{bavaud,ralf,kaufman}.  
Below we summarize the general structure of the method and apply it to interacting rigid molecules in equilibrium. The validity of
mean-field method is addressed in Appendices \ref{free_ap} and \ref{validity}. 

Consider $N$ identical interacting molecules with a potential energy
    \begin{equation}
            U(Q_1,...,Q_N)
        =\sum_{I=1}^N V_{\rm ext}(Q_I)+\frac{1}{2}\sum_{I\not= J=1}^N V(Q_I,Q_J),
\label{mok1}
    \end{equation}
where $V_{\rm ext}(Q_I)$ is the external potential, the inter-particle interaction $V(Q_I,Q_J)=V(Q_J,Q_I)$ is symmetric, and where $Q_I$ is the generalized coordinate of the particle $I$ to be specified below. The equilibrium Gibbs distribution for this system reads ($\beta=1/T$ is the inverse temperature) 
\be
\label{gibarian}
P(Q_1,...,Q_N)\propto e^{-\beta U(Q_1,...,Q_N)}.
\ee
Within the mean-field approach, partial distributions generated from $P(Q_1,...,Q_N)$ is replaced by $P(Q_1|Q_2)=P(Q_1)$, where the single-particle distribution $P(Q)$ reads 
\cite{vlasov,bavaud}:
    \begin{gather}
       \label{6} 
    P(Q)\propto e^{-\beta U_{\rm eff}(Q)}, \\     
    U_{\rm eff}(Q)=V_{\rm ext}(Q)+\frac{N-1}{Z}\int \d Q_1\, V(Q,Q_1) 
    e^{-\beta U_{\rm eff}(Q_1)},
\label{7}
    \end{gather}
where $U_{\rm eff}(Q)$ is the effective energy.
    
We apply (\ref{6}, \ref{7}) to $N$ charged molecules with a rigid charge distribution. All charges couple via the Coulomb interaction; $V_{\rm ext}$ refers to an external electrostatic potential energy. Each molecule is characterized by the central radius vector $\x$ and $n$ charges $\{q_i\}_{i=1}^n$ located at $\{\x+\a_i\}_{i=1}^n$. Hence, $Q=(\x,\aa)$, where $\aa=(\a_1,...,\a_n)$ and $\bb=(\b_1,...,\b_n)$. We get from (\ref{7}):
\be
\label{w1}
&&U_{\rm eff}(\x;\aa)=V_{\rm ext}(\x;\aa)+
\frac{N}{Z}\int_V \d \y\oint\d \bb\, e^{-\beta U_{\rm eff}(\y;\bb )} \sum_{i,j=1}^n \frac{q_iq_j }{|\x+\a_i-\y-\b_j|}
\\
\label{w2}
&&= V_{\rm ext}(\x;\a)+
\frac{N}{Z}\int_V \d \y\oint\d \b e^{-\beta U_{\rm eff}(\y;\b )} \T(\x,\aa)\T(\y,\bb)\frac{1}{ |\x-\y|},   \\
\label{w3}
&&\T(\x,\aa)=\sum_{j=1}^n q_je^{\a_j\na_{\x}}, \qquad \sum_{j=1}^n q_j=0,\\
\label{w4}
&&Z=\int_V \d \y\oint\d \bb \,e^{-\beta U_{\rm eff}(\y;\bb )}
\ee
where we replaced $N-1$ by $N$ assuming $N\gg 1$, introduced the shift operator $\T(\x,\aa)$, and 
and where $\oint\d\bb=\oint\d \b_1...\d \b_n$ means integration with all distances $|\b_i-\b_j|$ fixed (rigid body condition). The second equation in (\ref{w3}) means electric neutrality of the molecule. The integral $\int_V \d\y$ is taken over the volume $V$ of the system. The potential energy of the molecule due to the external field reads
\be
\label{v5}
V_{\rm ext}(\x;\aa)=\T(\x,\aa)\phi_{\rm ext}(\x),
\ee
where $\phi_{\rm ext}(\x)$ is the external potential. Hence the solution of (\ref{w2}) will be sought as 
\footnote{
Does (\ref{w5}) determine $\phi(\x)$ uniquely for a given $U_{\rm eff}(\x;\aa)$ and given $\T(\x,\aa)$? To answer this question, we look at solutions $\widetilde\phi(\x)$ of $\T(\x,\a)\widetilde\phi(\x)=0$, where a sufficiently smooth function $\widetilde\phi(\x)$ decays to zero for $|\x|\to \infty$. Hence, we can search for solutions via the Fourier expansion: $\widetilde\phi(\x)=\int \d^3 k\, e^{i\bm{k}\x}\widetilde\phi(\bm{k})$. This produces from (\ref{w3}): $\int \d^3 k\, e^{i\bm{k}\x}\widetilde\phi(\bm{k})\sum_{j=1}^n q_je^{i\a_j\bm{k}}=0$ for any $\x$. Hence, $\widetilde\phi(\bm{k})\sum_{j=1}^n q_je^{i\a_j\bm{k}}=0$ for any $\bm{k}$. For a sufficiently smooth $\widetilde\phi(\bm{k})$ this implies $\widetilde\phi(\bm{k})=0$. If $\widetilde\phi(\bm{k})=\sum_\ell\delta(\bm{k}-\bm{k}_\ell)$, then we get 
$\sum_{j=1}^n q_je^{i\a_j\bm{k}_\ell}=0$ for all $\bm{k}_\ell$. However, for such $\widetilde\phi(\bm{k})$, $\widetilde\phi(\bm{x})$ does not tend to zero for $|\x|\to\infty$. Thus, for the considered boundary conditions the only solution to $\T(\x,\a)\widetilde\phi(\x)=0$ is $\widetilde\phi(\x)=0$; i.e., (\ref{w5}) determines $\phi(\x)$ uniquely.
}
\be
\label{w5}
U_{\rm eff}(\x;\aa)=\T(\x,\aa)\phi(\x),
\ee
where the new unknown function (the self-consistent electric potential) is found from 
\be
\label{w6}
\phi(\x)=\phi_{\rm ext}(\x)+
\frac{N}{Z}\int_V \d \y\oint\d \bb\, e^{-\beta \T(\y,\bb)\phi(\y)} \T(\y,\bb)\frac{1}{ |\x-\y|}.
\ee
Applying to (\ref{w6}) the Laplace operator $\Delta$, and then integrating by parts (the boundary term nullifies due to $\delta(\x-\y)=\frac{1}{-4\pi}\Delta\frac{1}{|\x-\y|}$, once $\x$ is far from the boundary) we find from (\ref{w6}):
\be
\label{w7}
\Delta(\phi(\x)-\phi_{\rm ext}(\x))=-
\frac{4\pi N}{Z}\oint \d \bb\, \T(\x,-\bb)e^{-\beta \T(\x,\bb)\phi(\x)}.
\ee
This mean-field equation can also be obtained from the path-integral approach; cf.~\cite{buyukdagli}.

\section{Dimers in the nonlinear regime}
\label{section_dimers}

This section covers (finite-size) dimer molecules: the simplest scenario where nonlocality and nonlinearity play a significant role. 
Consider a dimer molecule: two charges $\pm q$ are fixed at the ends of vectors $\pm\a$. This implies in (\ref{w3})
\be
\label{dimo}
\T(\x,\aa)=q e^{\a\na_{\x}}-q e^{-\a\na_{\x}}.
\ee
Dimer molecules can be real (e.g., $C^{-0.07|q_{\rm e}|}\equiv O^{+0.07|q_{\rm e}|}$, where $q_{\rm e}<0$ is the electron charge, and where $-,\,=,\,\equiv$ denote covalent bonds) or represent colloidal particles; see Refs.~\cite{dimers_roads_review_klapp,dimers_roads_review_rutkow} for reviews on this interesting subject. Denoting for dimensionless quantities 
\be
\label{dimo}
\z=\x/|\a|,~~ \varphi=q\beta\phi, ~~ \varphi_{\rm ext}=q\beta\phi_{\rm ext}, 
\ee
we get for (\ref{w7}):
\be
\label{w87}
&&\Delta(\varphi(\z)-\varphi_{\rm ext}(\z))+\frac{d}{\pi}\,
\frac{ \oint \d \bm{e}\, [ e^{-\varphi(\z)+\varphi(\z-2\bm{e})}-
e^{-\varphi(\z+2\bm{e})+\varphi(\z)}  ] }{
1+\frac{1}{4\pi\hat{V}} \int_{\hat V} \d\z\oint \d \bm{e}\, [ e^{-\varphi(\z+\bm{e})+\varphi(\z-\bm{e})}-1]
}=0,\qquad |\bm{e}|=1,\\
&&
d=\frac{\pi N\beta q^2|\a|^2}{V}, \qquad \hat{V}=V/|\a|^3,
\label{ono}
\ee
where $\hat V\gg 1$ is the dimensionless volume, and where $\oint \d \bm{e}$ means integration over directions of the unit vector $\bm{e}$. Estimating
dimensionless $d$ via typical liquid concentration $N/V=10^{22}$
molecules$/{\rm cm}^3$, electron charge for $q$, room temperature 300 K,
and $|\a|=10^{-8}$ cm produces $d\sim 17$. In the point-like
(but generally nonlinear) limit, the equation (\ref{w87}) reproduces the results of
\cite{dpb1}. In the linearized case, it coincides with the
equation studied in \cite{buyukdagli,ralf}. 

The validity of the mean-field method is discussed in Appendices \ref{validity} and \ref{free_ap}, where we confirm that the mean-field free energy is an upper bound for the true free energy. The conclusion of Appendix \ref{validity} is that the mean-field method can have application range for $d\sim 1$ (our studies below will employ this regime), and it is certainly valid for $d\ll 1$. It is however nearly useless in the latter regime. 

To estimate the integral $\frac{1}{\hat{V}} \int \d\y\oint \d \bm{e}\, [ e^{-\varphi(\y+\bm{e})+\varphi(\y-\bm{e})}-1]$ in (\ref{w87}), we make several assumptions. All of them are confirmed below self-consistently. First, we assume that $\varphi(\y)$ is nowhere infinite. Otherwise, $e^{-\varphi(\y+\bm{e})+\varphi(\y-\bm{e})}$ will be infinite, and the integral will not converge. Third, we assume $\hat V\gg 1$, which is the thermodynamic limit. Fourth, we assume that due to external (non-point-like) charges located around $\y\sim 0$, the large-$|\y|$ behavior of $\varphi$ is $\varphi(\y)\sim |\y|^{-1}$.
Hence, in the dominant integration domain of the above integral we can apply:
\BEA
e^{-\varphi(\y+\bm{e})+\varphi(\y-\bm{e})}\simeq e^{-2\bm{e}\nabla\varphi(\y)}\simeq 1
-2\bm{e}\nabla\varphi(\y)+2[\bm{e}\nabla\varphi(\y)]^2.
\label{banakan}
\EEA
Using ($i,j=1,2,3$)
\BEA
\label{kensabanakan}
\oint \d \bm{e}=0,\qquad \oint\d \bm{e}\, e_ie_j=\frac{4\pi}{3}\delta_{ij},
\EEA
we find that the integral can be neglected in (\ref{w87}) due to
\be
\frac{1}{\hat{V}} \int \d\y\oint \d \bm{e}\, [ e^{-\varphi(\y+\bm{e})+\varphi(\y-\bm{e})}-1]={\cal O}(\hat V^{-1}).
\label{cardano}
\ee
We get from (\ref{w87}):
\be
\label{w88}
\Delta(\varphi(\z)-\varphi_{\rm ext}(\z))+\frac{2d}{\pi}\,
\oint \d \bm{e}\, \sinh\Big[-\varphi(\z)+\varphi(\z+2\bm{e})\Big]=0.
\ee
Our assumption of everywhere finite and localized $\varphi(\y)$ is consistent with assuming a spherically symmetric external fields: $\varphi_{\rm ext}(\z)=\varphi_{\rm ext}(r)$ ($r=|\z|$). It is then natural to assume in (\ref{w88}):
\BEA
\label{sphere}
\varphi(\z)=\varphi(r).
\EEA
Eq.~(\ref{sphere}) seems plausible, but it is an assumption. For mean-field (Poisson-Boltzmann) equations, it is known that spherically symmetric external fields
do not necessarily produce spherically symmetric solutions \cite{bavaud}. 

Now we can write (\ref{w88}) as
\be
\label{w89}
&&\Delta(\varphi-\varphi_{\rm ext})
+4d\int_{-1}^1\d \xi\, \sinh\Big[-\varphi(r)+\varphi(\sqrt{r^2+4+4r\xi})\Big]=0,\\
&&\Delta\varphi=\varphi''(r)+\frac{2}{r}\varphi'(r). 
\ee

\subsection{Linearization and permittivity}

The naive linearization of (\ref{w88}) refers to replacing $\sinh$ by
its argument.  This can be wrong in the vicinity of external sources
that generate $\varphi_{\rm ext}(\z)$. Assuming that these sources are
localized at $r=|\z|\sim 0$, we can linearize in the large-$r$ limit,
where the fields are anyhow weak. This will define the permittivity, which applies only in the large-$r$ limit. Expanding
$\varphi(\z+2\bm{e})-\varphi(\z)$ over $\bm{e}$ in (\ref{w88}) and using (\ref{kensabanakan}), we find
\be
\Delta(\varphi-\varphi_{\rm ext})+\frac{16d}{3}\Delta\varphi=0.
\label{ahn}
\ee
Eq.~(\ref{ahn}) means that the large-$r$ behavior of $\varphi$ is 
\be
\label{khan}
\varphi(r)=\frac{1}{1+\frac{16d}{3}}\varphi_{\rm ext}(r).
\ee
The expected message of (\ref{khan}) is that due to dipole screening, for large distances, the electric potential is smaller than the external one by a factor (permittivity) $\varepsilon=1+\frac{16d}{3}$. 

\begin{figure}[!htb]
\centering
  \subfigure[]{
   \includegraphics[width=0.475\columnwidth]{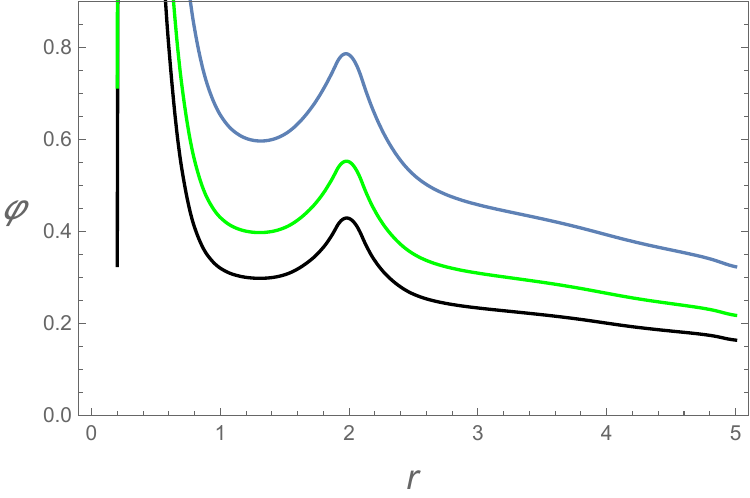} 
\label{fig01}
    } \hspace*{0.06cm}
    \subfigure[]{
    \includegraphics[width=0.475\columnwidth]{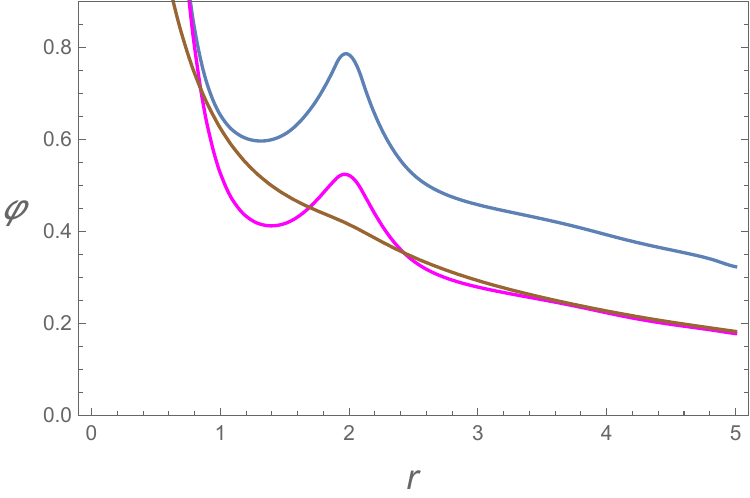}
    \label{fig02} 
    }
    \caption{ (a) Iterative solutions $\varphi(r)$ of (\ref{w89}) for dimers with $r_{\rm max}=5$ [cf.~(\ref{ontario2})], $b=9$, $c=50$ and various values of $d$: $d=5$ (blue), 
    $d=7.5$ (green), $d=11$ (\textcolor{blue}{black}). There is a partially visible maximum of $\varphi(r)$ at $\simeq 1/b$.
    \\ (b) The same as in Fig.~\ref{fig01}, but for $c=50$, $d=5$, and various values of $b$: $b=9$ (blue), $b=5$ (magenta), $b=1$ (brown).     }
\end{figure}

\comment{
\begin{figure}[ht]
\centering
\includegraphics[width=0.5\linewidth]{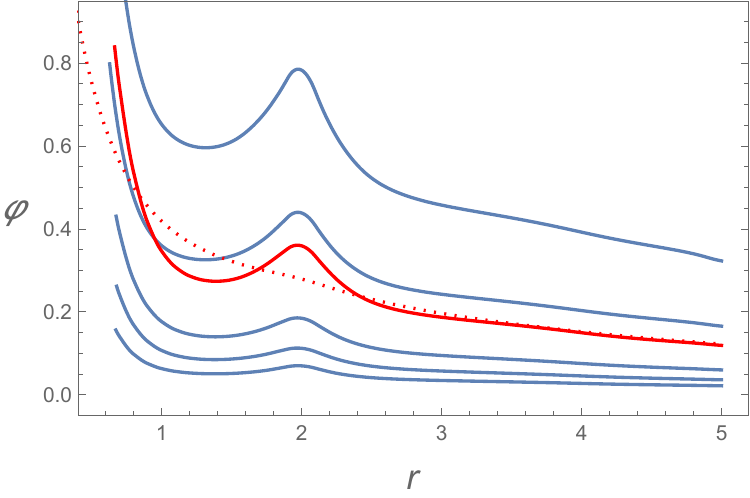}
\caption{Iterative solutions of (\ref{w89}) for dimers with $r_{\rm max}=5$ [cf.~(\ref{ontario2})] and convergence tolerance $10^{-4}$.  
Blue curves refer to the following parameters (from top to bottom): 
$b=9$, $c=50$ and $d=5$; 
$b=9$, $c=25$ and $d=5$; 
$b=5$, $c=25$ and $d=15$;
$b=5$, $c=15$ and $d=15$;
$b=5$, $c=15$ and $d=25$.
Red curve: $b=5$, $c=50$ and $d=15$. 
Red dotted curve: $b=1$, $c=50$ and $d=15$. 
It is seen that the potential minimum becomes deeper for larger $b$ and $c$ and smaller $d$. It disappears for $b\simeq 1$. 
}
 \label{fig0}
 \end{figure}
}

\subsection{Nonmonotonic mean-field potential and attraction of like charges }
\label{attra}

Now return to (\ref{w89}) and change there the variable:
\be
\label{w91}
&&(\chi-\chi_{\rm ext})''
+2d\int_{|r-2|}^{r+2}v\d v\, \sinh\Big[-r^{-1}\chi(r)+v^{-1}\chi(v)\Big]=0,\\
&& \chi(r)/r=\varphi(r), \qquad \chi(0)=0.
\label{chichkhan}
\ee

For solving (\ref{w91}) numerically, it is important that $[\chi_{\rm ext}(r)/r]|_{r\to 0}$ is finite. Due to this, also $[\chi(r)/r]|_{r\to 0}$ is finite, which was already assumed before (\ref{cardano}); see our assumptions before (\ref{banakan}). Hence, for the localized external field, we choose the following model of an extended (non-point-like) charge
\be
\label{amba}
\phi_{\rm ext}(\x)=\frac{Q(1-e^{-|\x|/r_1}) }{|\x|},
\ee
where $Q$ is the total charge, and $r_1$ is the range over which this external charge is localized. Eq.~(\ref{amba}) reads in dimensionless units [see (\ref{dimo}, \ref{chichkhan})]: $\phi_{\rm ext}=\chi_{\rm ext}/|\z|$,
\be
\label{ugol}
\chi_{\rm ext}=c(1-e^{-b r}), \quad c=\frac{Q q\beta}{|\bm{a}|},\quad 
b=\frac{|\bm{a}|}{r_1}.
\ee
In particular, (\ref{ugol}) can describe ions \footnote{Typical parameters here are as follows. $r_1(e^-)$ and $r_1(H^+)$ are much smaller than $\AA=10^{-8}$ cm. Positive ions are typically smaller than negative ones due to less electron repulsion: 
$r_1(Li^{+})=0.76 \AA$, $r_1(Na^+)=1.02 \AA$, $r_1(Cl^-)=1.81 \AA$. 
}.

Eq.~(\ref{w91}) is an integro-differential equation, which contains both advanced and retarded arguments. Solving such equations is not straightforward. We solved it numerically and iteratively within a convex set of functions that hold 
\be
\label{ontario1}
&\chi(0)=0,\\ 
\label{ontario2}
&\chi(r\geq r_{\rm max})= c/(1+\frac{16d}{3}),\\ 
&\chi'(r=r_{\rm max})=0,
\label{ontario3}
\ee
where $r_{\rm max}$ is sufficiently large. Condition (\ref{ontario1}) comes from $\varphi(r)=\chi(r)/r$ and finiteness of $\varphi(0)$, which relates to the finiteness of $\varphi_{\rm ext}(0)$. Condition (\ref{ontario2}) comes from (\ref{khan}). This condition is needed because solving (\ref{w91}) in any finite interval $r\in(0,r_{\rm max}]$ will require inputs $r\in(0,r_{\rm max}+2]$. Eq.~(\ref{ontario3}) comes from the smoothness. The actual solution is easier after integrating (\ref{w91}) once:
\be
\label{w94}
(\chi-\chi_{\rm ext})'
+2d\int_{r_{\rm max}}^r\d u \int_{|u-2|}^{u+2}v\d v\, \sinh\Big[-u^{-1}\chi(u)+v^{-1}\chi(v)\Big]+
\chi_{\rm ext}'(r_{\rm max})
=0,
\ee
where the integration constant was chosen from requiring $\chi'(r_{\rm max})=0$.
Eq.~(\ref{w94}) is solved iteratively. Note that the integro-differential equations with both advanced and retarded arguments are not guaranteed to have always well-defined solutions; see e.g. \cite{driver}. In fact, iterative methods is generically the only way to give meaning to such equations.  

Figs.~\ref{fig01} and \ref{fig02} show numerical solutions of (\ref{w94}). The most
interesting aspect of this solution is that $\varphi(r)=\chi(r)/r$ is a
nonmonotonous function of $r$ for a sufficiently large $b$ and $c$ in
(\ref{ugol}), i.e., for a sufficiently localized and strong external
charge. This means that a test charge $Q_1$ that has the same sign as the
external charge $Q$ can be bound to it; cf.~(\ref{amba}). Note that we need $|Q_1|\ll |Q|$, so that the test particle sees $Q_1\phi$ as an external potential. 

The barrier separating the local minimum of $\varphi$ is located at the dimensionless distance $r=2$, which is precisely the dimer (dimensionless) length; see Figs.~\ref{fig01} and \ref{fig02}. This suggests the local minimum arises due to dimers that are bound to the external charge and screen this charge. Note from Fig.~\ref{fig02} that the local minimum exists for a sufficiently bounded external charge (i.e., a large $b$ in (\ref{ugol})), and also that this minimum becomes shallow for larger
values of $d$ in (\ref{w94}); see Fig.~\ref{fig01}. The minimum is best pronounced for intermediate values of $d$ (see Fig.~\ref{fig01}), because it is clearly absent for $d\to 0$, where $\varphi\to\varphi_{\rm ext}$. 

The main known mechanism for the attraction of like charges relates to
counterion correlations between charges, macroions, which have a much larger size
than the molecular scale \cite{levin,ali}. This effect is essentially
non-mean-field \cite{levin,ali}, governed by the strong coupling parameter $\Xi=2\pi q^3\ell_B^2\sigma$, where $q$ is the valence of counterions, $\ell_B$ is the Bjorrum length, and $\sigma$ is the surface charge density of macroion. Similarly and highly charged surfaces in the presence of multivalent counter-ions attract each other in the strong coupling regime $\Xi > 1$, in contrast to the Poisson-Boltzmann limit, where the interaction is always repulsive. In our situation, the attraction
between (effectively point-like) charges is a mean-field effect, but it
is driven by correlations present due to the non-point-like dipole
structure. As shown in Figs.~\ref{fig01} and \ref{fig02}, the existence of the bound state is primarily governed by the dielectric permittivity $\varepsilon=1+\frac{16d}{3}$ [cf.~(\ref{khan})], and the localization of external charge. According to (\ref{khan}), the $d$ parameter increases linearly with the dielectric permittivity $\varepsilon$, which results in the shallower minima in Fig.~\ref{fig01}. The external charge localization is regulated by the $b$ parameter, which results in the appearance of the bound state between the external and test charges in Fig.~\ref{fig02}.

\comment{ 
%results by Alex and Varazdat show that more elaborated algorithms 
% do not show convergence problems

\subsection{Convergence problem of iterations}
\label{conve}

In the context of iterative numerical solutions of (\ref{w94}), we
should note that they have a clear convergence problem, if the medium is
sufficiently nonlinear, i.e. $d$ is large, while the external field is
sufficiently strong, i.e. $b$ and $c$ in (\ref{ugol}) are large. For
example, independently from the initial iteration point and the
convergence tolerance, the iterations did not converge at all for $b=9$,
$c=50$ and $d=15$, while for $b=9$, $c=50$ and $d=10$ the convergence
tolerance was never smaller than $10^{-2}$, instead of $10^{-4}$ applied
in those cases, where we claim convergence. This lack of convergence may
indicate on the absence of solutions, since the integro-differential
equations with both advanced and retarded arguments are not guaranteed
to have always well-defined solutions; see e.g. \cite{driver}. In fact, iterative
methods is generically the only way to give meaning to such equations.  

More research is needed to understand this lack of convergence; recall 
our discussion after (\ref{sphere}). In this
context, we stress that solutions of the linearized equation (\ref{w94})
are always well-defined and can be found via the Fourier transform. The
linearization holds for a small $d$.  Also, for point-like dipoles, the
resulting nonlinear equation is well-defined for all $d$ \cite{dpb1}.
The point-like limit assumes $|\a|\to 0$. 
}

\begin{figure}[ht]
\centering
\includegraphics[width=0.6\linewidth]{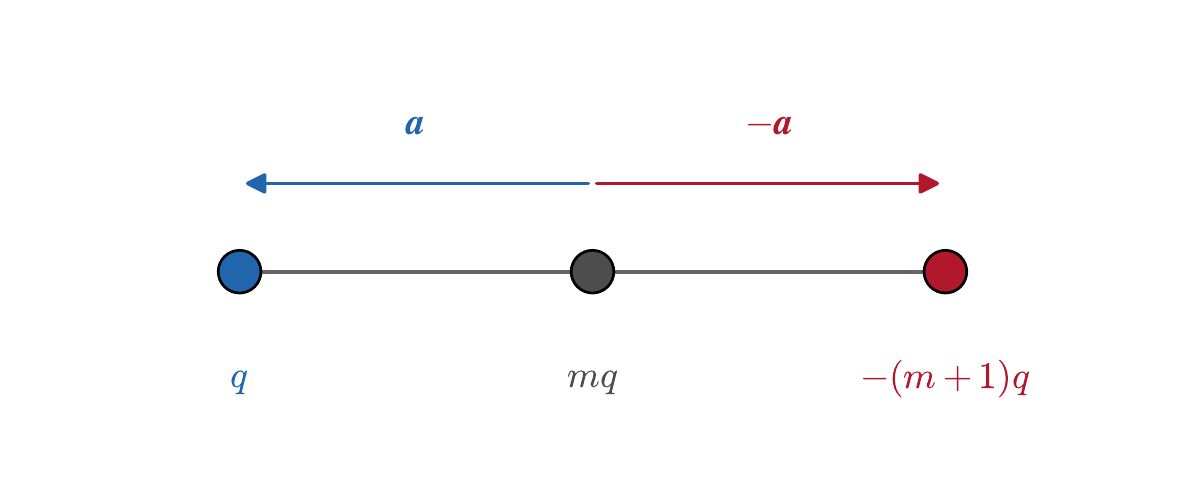}
\caption{Schematic representation of a linear molecule with charges $q$, $mq$, and $-(m+1)q$; see (\ref{dom}). 
}
 \label{fig00}
 \end{figure}

\section{The effective C-symmetry (${\rm C}_{\rm eff}$-symmetry) violation}
\label{section_generalized_dimers}

\subsection{Generalized dimers: linear molecules in the nonlinear situation}

In this section, we study linear molecules, where the charge distribution is more complex than with dimers. Consider a linear molecule [cf.~(\ref{dimo}) and see Fig.~\ref{fig00}], where 
\be
\label{dom}
\T(\x,\a)=q e^{\a\na_{\x}}+mq e^{-\a\na_{\x}}-(m+1)q,
\ee
i.e., charges $q$ and $mq$ are located at $\a$ and $-\a$, respectively, while the charge $-(m+1)$ is located at $0$. Eq.~(\ref{dom}) describes linear (generally asymmetric) molecules; e.g., the nitrosyl fluoride molecule ($F-N=O$) can be modeled via 3 point-like charges arranged along a line \cite{hirshfeld}: $F^{-0.25|q_{\rm e}|}-N^{0.4|q_{\rm e}|}=O^{-0.15|q_{\rm e}|}$, where we indicate partial charges around each atom in terms of the electron charge $q_{\rm e}<0$. We now get instead of (\ref{w88}) \footnote{The reasoning around (\ref{banakan}--\ref{cardano}) is now slightly different, but we still reach the conclusion after (\ref{cardano}) about neglecting the integral.}
\be
\label{w888}
\Delta(\varphi(\x)-\varphi_{\rm ext}(\x))+\frac{d}{\pi}\,
\oint \d \bm{e}\,\Big[ e^{(m+1)\varphi(x+\bm{e})-\varphi(\x)-m\varphi(\x+2\bm{e})}
+me^{(m+1)\varphi(x+\bm{e})-\varphi(\x+2\bm{e})-m\varphi(\x)}\nonumber\\
-(m+1)e^{(m+1)\varphi(x)-\varphi(\x+\bm{e})-m\varphi(\x-\bm{e})}
\Big]
=0.
\ee
We generalize (\ref{khan}) by expanding the nonlinear terms in (\ref{w888}) over $\varphi$, and 
using (\ref{kensabanakan}):
\be
\label{khan2}
\varphi(r)=\frac{1}{1+\frac{4d((m+1)^2-4m)}{3}}\varphi_{\rm ext}(r).
\ee
Eventually, (\ref{w94}) is generalized by replacing there $\sinh\Big[-u^{-1}\chi(u)+v^{-1}\chi(v)\Big]$ with:
\be
&&\frac{1}{2}\exp\Big[(m+1)\frac{\chi(\kappa_1)}{\kappa_1}-\frac{\chi(u)}{u}-\frac{m\chi(v)}{v}\Big]
+\frac{m}{2}\exp\Big[(m+1)\frac{\chi(\kappa_1)}{\kappa_1}-\frac{m\chi(u)}{u}-\frac{\chi(v)}{v}\Big]\nonumber\\
&&-\frac{m+1}{2}\exp\Big[-\frac{\chi(\kappa_1)}{\kappa_1}+\frac{(m+1)\chi(u)}{u}
-m\frac{\chi(\kappa_3)}{\kappa_3}
\Big],\nonumber\\
&& \kappa_1=\sqrt{\frac{u^2}{2}  +\frac{v^2}{2}-1  }, \qquad
\kappa_3=\sqrt{\frac{3u^2}{2}  -\frac{v^2}{2}+3  }.
\label{w944}
\ee
Eqs.~(\ref{ontario1}, \ref{ontario3}) stay intact, but (\ref{ontario2}) should be replaced with $\chi(r\geq r_{\rm max})= c/(1+  \frac{4d((m+1)^2-4m)}{3} )$; cf.~(\ref{khan2}).

\comment{
\begin{figure}[ht]
                \centering
                \includegraphics[width=0.5\linewidth]{figure1.pdf}
                \caption{For the linear molecule (\ref{dom}) with $m=-1/2$, we plot the solution $\chi=\chi_+$ of dimensionless (\ref{w888}, \ref{ugol}) under $c=5$, $d=5$ and $b=5$ and $r_{\rm max}=5$. This is the red curve, where $r_{\rm max}$ is defined in (\ref{ontario2}). We also plot the solution $-\chi=-\chi_-$ of (\ref{w888}, \ref{ugol}) under $c=-5$, $d=5$ and $b=5$ and $r_{\rm max}=5$ (blue curve). If the symmetry under (\ref{symsym}) were correct, we would have $\chi_+=-\chi_-$. The actual deviation between $\chi_+$ and $-\chi_-$ confirms that this symmetry is broken. Note that no symmetry-breaking effects are visible for $|c|=1$, $d=1$, and $b=1$, where we have $\chi_+\approx -\chi_-$ with a good approximation. }
                \label{fig1}
            \end{figure}
}

\begin{figure}[ht]
                \centering
                \includegraphics[width=0.5\linewidth]{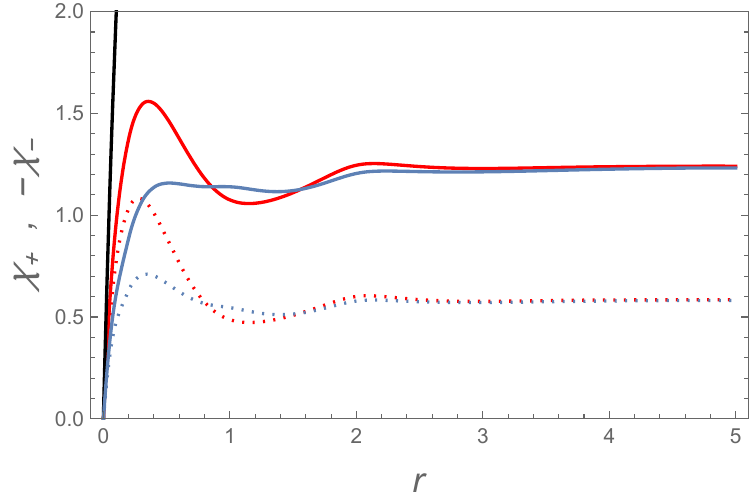}
                \caption{ For the linear molecule (\ref{dom}) with $m=-1/2$, we plot the solution $\chi=\chi_+$ of dimensionless (\ref{w888}, \ref{ugol}) under $c=5$, $d=1$ and $b=5$ and $r_{\rm max}=5$. This is the red curve, where $r_{\rm max}$ is defined in (\ref{ontario2}). We also plot the solution $-\chi=-\chi_-$ of (\ref{w888}, \ref{ugol}) under $c=-5$, $d=1$ and $b=5$ and $r_{\rm max}=5$ (blue curve); $c=5$, $d=2.5$ and $b=5$ (red dotted); $c=-5$, $d=2.5$ and $b=5$ (blue dotted). The black curve denotes $\chi_{\rm ext}(r)=c(1-e^{-br})$ for $c=5$ and $b=5$; cf.~(\ref{ugol}). Note that for small $r$ the black curve nearly coincides with red and blue curves, which indicates that (naturally) the screening is absent at small distances.
                }
                \label{fig2}
            \end{figure}

\subsection{${\rm C}_{\rm eff}$-symmetry }

Now compare (\ref{w88}) with (\ref{w888}). Eq.~(\ref{w88}) is invariant with respect to 
\BEA
\label{symsym}
\varphi_{\rm ext}\to -\varphi_{\rm ext}, \quad \varphi\to -\varphi.
\EEA
This symmetry means that if one keeps the medium intact, but changes the sign of the external charge that produces $\varphi_{\rm ext}$, this new charge will produce the same potential but with the inverted sign (i.e., $-\varphi$). Now the linearized mean-field situation is invariant under (\ref{symsym}), but (\ref{w88}) stays invariant also in the nonlinear situation, when $\sinh[x]\not=x$ in (\ref{w88}). In contrast, the nonlinear situation of (\ref{w888}) is not invariant under (\ref{symsym}), as checked directly. Note that (\ref{symsym}) is not the full charge-inversion, which of course keeps invariant both (\ref{w88}) and (\ref{w888}), as well as any other electrodynamic relation. Eqs.~(\ref{w88}, \ref{w888}) hold this full symmetry, since due to the definition of $\varphi_{\rm ext}$ and $\varphi$ [see after (\ref{dimo})], they both stay invariant under inversion of all the charges. Eq.~(\ref{symsym}) refers to an effective charge-conjugation, where the external charges change their sign, but they are immersed in the same medium. 

Importantly, also the nonlinear, but local version of (\ref{w7}) is invariant under (\ref{symsym}). This nonlinear equation is obtained e.g. from (\ref{w7}) and (\ref{dimo}) by (formally) expanding (\ref{w7}) over ${\bm e}$. It has the usual electrodynamical form \cite{landau8}:
\BEA
\label{yula}
\nabla \Big(\varepsilon(|\nabla \phi|) \nabla \phi\Big)=-4\pi\rho_{\rm ext}=\Delta 
\phi_{\rm ext}.
\EEA
Its symmetry under (\ref{symsym}) follows from the fact the electric
permittivity $\varepsilon(|\nabla \phi|)$ (in the considered isotropic
system) depends on the absolute value of the electric field $\nabla
\phi$ only \cite{landau8}. For this argument, we do not need the explicit
form of $\varepsilon(|\nabla \phi|)$, though this form can be written
down. This isotropy argument can be extended to, for example, quadrupole
systems. Note, however, that in the nonlinear domain, the argument requires that spatial nonlocality be absent. Otherwise, also space derivatives can enter the game,
and the simple form of $\varepsilon(|\nabla \phi|)$ in (\ref{yula}) is lost.

In Fig.~\ref{fig2}, we employed the following parametrization, which is also applied to all other figures. We assume that the external field (\ref{ugol}) can be generated by a charge having $Q>0$ [red curve] or $Q<0$ [blue curve], i.e., $c>0$ and $c<0$, respectively; cf.~(\ref{ugol}). The molecular charge $q$ in (\ref{dom}) (see also Fig.~\ref{fig00}) is kept always positive $q>0$. Hence, both curves in Fig.~\ref{fig2} refers to molecules $(q,-0.5q,-0.5q)$.

The main message of Fig.~\ref{fig2} is that solutions of (\ref{w944}, \ref{ugol}) are indeed not invariant under $c\to -c$ and $\chi\to-\chi$, i.e., under (\ref{symsym}). If the symmetry under (\ref{symsym}) were correct, we would have $\chi_+\equiv\chi(c>0)=-\chi_-\equiv -\chi(-c<0)$, which is not the case in Fig.~\ref{fig2}. It is seen that the negative charge is screened more, since $\chi_+> -\chi_-$ holds for most values of $r$. The reason for this can be explained as follows. Obviously, the negative (positive) external charge is screened by positive (negative) charges in $(q,-0.5q,-0.5q)$ ($q>0$). Now note that the charge density in $(q,-0.5q,-0.5q)$ is larger for the positive charge $q$. Note as well that the screening itself means $|\chi_+(r)|< |\chi_{\rm ext}(r)|$ and $|\chi_-(r)|<|\chi_{\rm ext}(r)|$, which is also seen in Fig.~\ref{fig2}. 

How does the ${\rm C}_{\rm eff}$-symmetry violation depend on the involved parameters? 
Fig.~\ref{fig2} shows that the difference $\chi_+-|\chi_-|$ is not especially susceptible to decreasing $d$ (with $c$ and $b$ being fixed). As expected, for larger $d$ both $\chi_+$ and $|\chi_-|$ are smaller: a larger molecular concentration leads to stronger screening. However, the difference $\chi_+-|\chi_-|$ crucially depends on $c$ and $b$: the difference $\chi_+-|\chi_-|$ disappears for smaller $c$ (less charged external field) and for smaller $b$ (more broadly distributed external charge). For example, under parameters for $|c|=1$, $d=1$, and $b=1$ [cf.~Fig.~\ref{fig2}], $\chi_+$ and $-\chi_-$ nearly coincide with each other. 

Recall that $r=1$ is the (dimensionless) distance between each partial charge in $(q,-0.5q,-0.5q)$. In Fig.~\ref{fig2}, we see that the first maximum of both $\chi_+$ and $-\chi_-$ occurs around $r=1/b=0.2$, which is the effective radius of the external charge; see (\ref{ugol}). The second maximum for both curves occurs around $r=2$; it coincides with the overall size of the molecule. These aspects reflect the structure of the solvation layer around the external charge.  

\subsection{Free energy and entropy}
\label{free}

Here we shall describe the above ${\rm C}_{\rm eff}$-symmetry violation via thermodynamic quantities. 
For the mean energy of the considered mean-field electrostatic system, we have [see Appendix \ref{ella}]:
\be
\bar{U}
=\frac{1}{8\pi}\int \d \y \, [(\na\phi(\y))^2-(\na\phi_{\rm ext}(\y))^2].
\label{ww9}
\ee
The meaning of (\ref{ww9}) is that this is the full electric energy minus the energy of the external field. Note that (\ref{ww9}) applies to the nonlinear situation. Likewise, we have for the free energy from \footnote{We do not include in (\ref{ww10}) the factor $\ln[N!]$, since we do not consider situations where the number of particles changes.}:
\be
F=-TN\ln Z-\frac{1}{8\pi}\int \d \y \, [\na\phi(\y)-\na\phi_{\rm ext}(\y)]^2.
\label{ww10}
\ee
Recall that the free energy (\ref{ww10}) has two related meanings. First, $F$ is an upper bound to the true free energy (true means not employing the mean-field approach). Second, (\ref{ww10}) describes the isothermal work done upon slowly changing an external parameter of the system; see Appendix \ref{free_ap}.

Eq.~(\ref{ww9}) is the basis for Born's solvation model, which is a known phenomenological approach; see \cite{born_model} for a recent review of this model \footnote{Because Born's model uses the homogeneous dielectric constant as an input parameter, it does not apply for those cases (such as our case) where this constant is ill-defined at short distances due to nonlocal and nonlinear effects. An unclear point in Born's model is why the phenomenological reasoning is applied to energy (\ref{ww9}) rather than directly to the free energy (\ref{ww10}), which is the potential function of the canonical ensemble. A related problem is discussed in \cite{born_model}. }. We go beyond such approaches, since we have the well-defined free energy (\ref{ww10}), and we already calculated $\phi$ numerically. 

Recall from (\ref{w4}, \ref{cardano}) that $-TN\ln Z$ amounts to an additive factor, since (for large volumes) we approximate $Z= V\oint\d \bb$. The factor $-TN\ln Z$ in (\ref{ww10}) is not relevant for our consideration, since below we focus on free energy and entropy differences, where it cancels out. We now apply (\ref{ww9}, \ref{ww10}) to the solution of (\ref{w91}--\ref{ontario3}):
\be
\label{asala1}
F&=&-TN\ln Z-\frac{T^2|\a|}{2q^2}
\int_0^\infty \d r \Big[\chi'-\chi'_{\rm ext}-\frac{\chi}{r}+\frac{\chi_{\rm ext}}{r}\Big]^2,\\
S&=&N\ln Z+\frac{T|\a|}{q^2}\int_0^\infty \d r \Big[\chi'-\chi'_{\rm ext}-\frac{\chi}{r}+\frac{\chi_{\rm ext}}{r}\Big]^2\\
\label{asala2}
&+&\frac{T|\a|}{q^2}\int_0^\infty \d r \Big[\chi'-\frac{\chi}{r}\Big]
\Big[\chi'_{\rm ext}-\frac{\chi_{\rm ext}}{r}\Big],
\label{asala3}
\ee
where for the entropy $S$ we employed $TS=\bar{U}-F$. The integrals in (\ref{asala1}--\ref{asala3}) can be separated as $\int_0^\infty=\int_0^{r_{\rm max}}+\int_{r_{\rm max}}^\infty$. Now only the $\int_0^{r_{\rm max}}$ parts contribute to the free-energy and entropy differences that are due to the sign-difference of $\chi_{\rm ext}$. Indeed, in the $\int_{r_{\rm max}}^\infty$ parts 
we can take $\chi'=\chi_{\rm ext}'=0$ and $\chi\propto\chi_{\rm ext}$, i.e., those parts
depends only on $\chi_{\rm ext}^2$ and $\varepsilon=1+\frac{4d((m+1)^2-4m)}{3}$; see (\ref{khan2}). 

We now apply these formulas to the situation shown in Ref.~\ref{fig2}. Denoting by $F_{\pm}$ and $S_{\pm}$ the free energy and entropy of $\chi_\pm$ solutions, respectively, we get for parameters 
$c=\pm 5$, $d=1$ and $b=5$ of Fig.~\ref{fig2}:
\be
F_+-F_-=8.5458\, \frac{T^2|\a|}{2q^2}, \qquad 
S_+-S_-=-0.704\, \frac{T|\a|}{q^2}.
\label{umekawa}
\ee
The first message of (\ref{umekawa}) is that the concept of better screening, which we identified via a smaller absolute value of $\varphi$ (or $\chi$), has a clear thermodynamic meaning, since it relates to smaller free energy. Recall from Appendix \ref{free_ap} that the mean-field free energy (\ref{ww10}) is not only a variational upper bound for the true free energy, but is also an isothermal work done when slowly changing an external parameter.
Hence, the lower free energy in (\ref{umekawa}) means that when the external charge is slowly changed, the work is extracted when going to better screening. 

The second message of (\ref{umekawa}) is that better screening relates to a larger entropy.
Though less expected, this information is pertinent to understanding the screening structure. 

\begin{figure}[ht]
                \centering
                \includegraphics[width=0.5\linewidth]{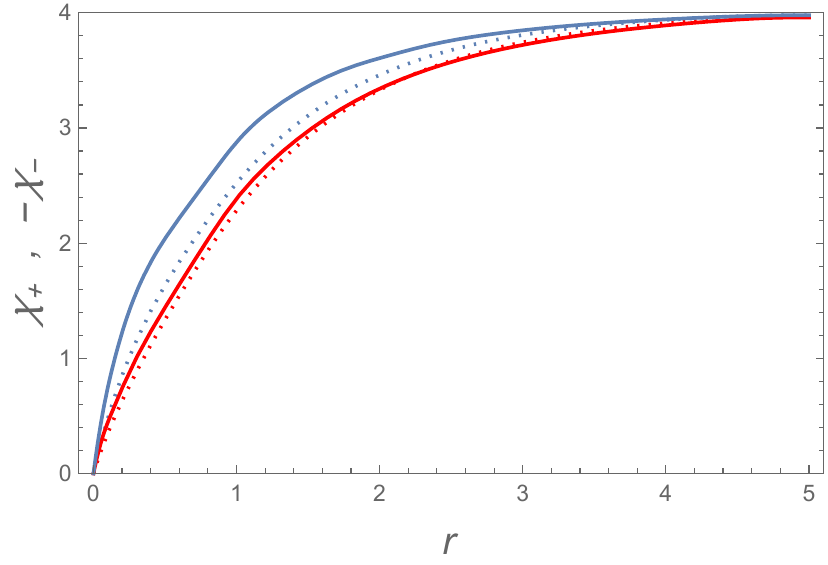}
                \caption{ For the linear molecule (\ref{dom}) with $m=1$, we plot the solution $\chi=\chi_+$ of dimensionless (\ref{w888}, \ref{ugol}) under $c=4$, $d=1$ and $b=4$ and $r_{\rm max}=5$. This is the red curve, where $r_{\rm max}$ is defined in (\ref{ontario2}). We also plot the solution $-\chi=-\chi_-$ of (\ref{w888}, \ref{ugol}) under $c=-4$, $d=1$ and $b=4$ and $r_{\rm max}=5$ (blue curve). Dotted curves denote $\chi_+$ and $-\chi_-$ for $c=\pm 4$, $d=1$ and $b=2$ and $r_{\rm max}=5$; i.e., for a smaller value of $b$, where $\chi_+-|\chi_-|$ is smaller. 
                }
                \label{fig22}
            \end{figure}

\comment{ {\bf These numerical results show that the entropy difference changes its sign upon decreasing $b$. I do not generally trust it, because the numbers are small and more numerical precision is needed.}

A general conclusion about the free energy is that more screening is associated with a lower free energy. But the entropy differences depend on the involved parameter regime. For example, while (\ref{umekawa}) refers to parameters $c=\pm 5$, $d=1$, $b=5$, and $m=-0.5$ of Fig.~\ref{fig22}, we get for a smaller value $b=2$ (instead of $b=5$) with all other parameters being fixed:
\be
F_+-F_-=0.6630\, \frac{T^2|\a|}{2q^2}, \qquad 
S_+-S_-=0.0547\, \frac{T|\a|}{q^2}.
\label{umekawa2}
\ee
For this smaller value of $b$, i.e., for a more smeared external charge, we (as expected) have that the difference between $\chi_+$ and $\chi_-$ is less pronounced; hence $F_+-F_-$ in (\ref{umekawa2}) is positive, but smaller than in (\ref{umekawa}). However, the entropy difference in (\ref{umekawa2}) changed its sign compared to (\ref{umekawa}).
}

\subsection{Linear symmetric molecule with zero dipole moment}

Fig.~\ref{fig2} studied a linear molecular structure $(q,-0.5 q, -0.5q)$ [cf.~(\ref{dom})], which is similar to a dimer with one of its ends having a lower charge density. A linear structure with zero dipole momentum that is possibly far from the dimer refers to $m=1$ in (\ref{dom}); i.e., it refers to the linear molecule $(q,-2q,q)$ ($q>0$). Notable examples of such molecules are carbon dioxide $O^{-0.3|q_{\rm e}|}=C^{0.6|q_{\rm e}|}=O^{-0.3|q_{\rm e}|}$, and carbon disulfide $S^{-0.18|q_{\rm e}|}=C^{0.36|q_{\rm e}|}=S^{-0.18|q_{\rm e}|}$. 

Consider now the situation where the external field is given by (\ref{ugol}). We pose the same question as above: which external charge $Q>0$ ($c>0$) or $Q<0$ ($c<0$) is screened better. The intuition drawn from the previous example (see 
Fig.~\ref{fig2}) implies that $Q>0$ will be screened better, because the absolute value of the negative charge density in $(q,-2q,q)$ is larger than the positive charge density, i.e., the negative charge is locally stronger than the positive charges. Fig.~\ref{fig22} shows that this intuition is incorrect: it is $Q<0$ that is screened better. One possible reason for this is that 
different molecules $(q,-2q,q)$ that should screen the positive external charge perturb each other via their positively charged end-points. 

The message of Fig.~\ref{fig22} can be confirmed thermodynamically using (\ref{asala1}--\ref{asala3}):
\be
F_+-F_-=-7.4599\, \frac{T^2|\a|}{2q^2}, \qquad 
S_+-S_-=1.2951\, \frac{T|\a|}{q^2}.
\label{umekawa3}
\ee
We see that $\chi_+$ has less free energy (better screening) and also larger entropy. In that respect (less free energy and more entropy) (\ref{umekawa3}) and (\ref{umekawa}) are similar. 

Fig.~\ref{fig22} also shows $\chi_+$ and $-\chi_-$ for smaller value of $b$ ($b=2$). It is seen that the asymmetric screening is much smaller, but still visible: the free energy difference for this case amounts to $F_+-F_-=-1.7576\, \frac{T^2|\a|}{2q^2}$ instead of (\ref{umekawa3}). 

\begin{figure}[ht]
                \centering
                \includegraphics[width=0.5\linewidth]{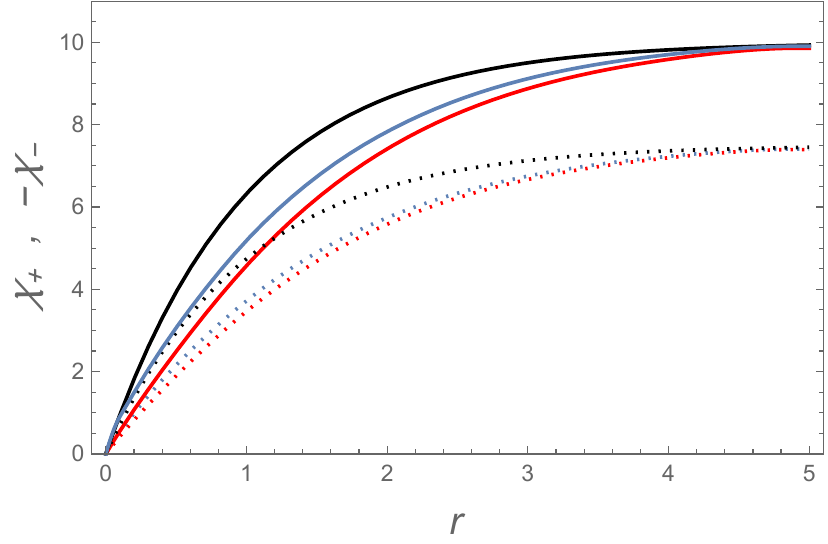}
                \caption{ The same as in Fig.~\ref{fig22}, but with parameters $m=1$, $b=1$ and $d=1$. Red curve $\chi_+$ for $c=10$; blue curve $c=-10$; dotted red $c=7.5$; dotted blue $c=-7.5$. Black and dotted black curves shows, respectively, $\chi_{\rm ext}=c(1-e^{-b r})$ for $c=10$ and $c=7.5$. The (dotted) red and blue curves are below the (dotted) black curve, which illustrates the short-distance screening effect. 
                }
                \label{fig33}
            \end{figure}

Fig.~\ref{fig33} looks at a different regime: now $b=1$ means that the characteristic radius of the external charge is of the same order as the molecular length $|\a|$; cf.~(\ref{ugol}). Now the asymmetry effect is visible for a larger $c=10$ (i.e., for a stronger charging), and it is almost invisible for $c=7.5$. 
Fig.~\ref{fig33} also illustrates the sense in which the screening is present. Note from (\ref{khan2}) that for the considered zero dipole moment situation ($m=1$), the large-$r$ values of $\chi(r)$ and $\chi_{\rm ext}(r)$ are the same. Nevertheless, we have $\chi_{\rm ext}(r)\geq\chi(r)$, as expected from the short-distance screening; see Fig.~\ref{fig33}. 

The behavior of $\chi_+$ and $\chi_-$ under changing $d$ (with other parameters being fixed) is as follows (not shown in figures). Both $\chi_+$ and $|\chi_-|$ decrease under increasing $d$, since now the density of molecules (and hence screening) are larger; cf.~(\ref{ono}). Simultaneously, $\chi_+-|\chi_-|$ is smaller for larger values of $d$.

Hence, when linear molecules with a zero dipolar momentum have to screen an external charge, those charges are screened better which attract the end-points of the molecule. This conclusion coincides with the results on asymmetric screening obtained for water molecules within phenomenological and molecular dynamics approaches \cite{asymmetric_solvation_0,asymmetric_solvation_1,asymmetric_solvation_2,asymmetric_solvation_3,asymmetric_solvation_4}. In the present work we deliberately refrain from studying water molecules, also because there are legitimate doubts about applying the mean-field method to this protic liquid. 

\comment{
Speculatively, we can nevertheless attempt to explain the above coincidence as follows. The water molecule can be qualitatively represented as a combination of the linear molecule with zero dipole momentum and a symmetrical dimer. Once the for the dimer the asymmetry in screening is absent due to symmetry (even in the nonlinear and nonlocal regime), the asymmetric screening for the water molecule qualitatively agrees with that of the zero dipole moment molecules. We emphasize that this explanation is speculative and should serve as a starting point for more elaborate approaches.
}

\begin{figure}[ht]
                \centering
                \includegraphics[width=0.5\linewidth]{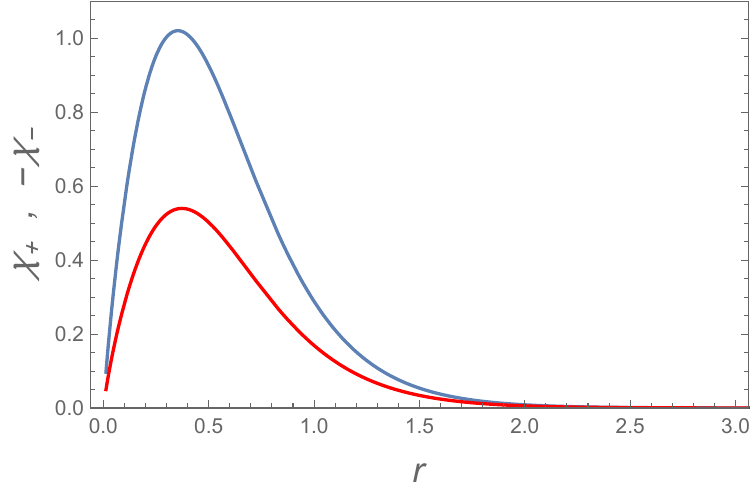}
                \caption{ For the asymmetric electrolyte described by (\ref{poi2}, \ref{poi22}, \ref{ugol}), we plot the solution $\chi=\chi_+$ under $c=10.$, $a^2=15.$ and $b=5$. This is the red curve. We also plot $-\chi=-\chi_-$ (blue curve), where $\chi_-$ solves (\ref{poi2}, \ref{poi22}) under $c=-10$, $a^2=15$, and $b=5$. Note that $\phi(r)=\chi(r)/r$ is monotonically decaying for both curves.
                }
                \label{fig3}
            \end{figure}

\section{${\rm C}_{\rm eff}$-symmetry violation for the nonlinear Poisson-Boltzmann systems}
\label{poisson}

Above we studied the ${\rm C}_{\rm eff}$-symmetry violation for two types of linear molecules 
$(q,-0.5 q, -0.5q)$ and $(q,-2q, q)$. To gain deeper understanding of more general scenarios
of the ${\rm C}_{\rm eff}$-symmetry violation, we shall focus on 
asymmetric electrolytes, where the intermolecular bonds are broken. For definiteness, we choose calcium chloride, ${\rm Ca}^{+2}+2{\rm Cl}^{-}$, which electroneutral, but contains ions with different charges.
The mean-field theory of such electrolytes is given by the Poisson-Boltzmann equation
(see, e.g., \cite{tellez_review,gray_review,xing} for reviews):
\be
\label{poi1}
\Delta(\phi(\x)-\phi_{\rm ext}(\x))+
\frac{4\pi}{V}\Big[ N_1q_1 e^{-\beta q_1\phi}  +N_2q_2 e^{-\beta q_2\phi}  \Big]=0,
\ee
where $\phi$ and $\phi_{\rm ext}$ are (resp.) the full mean-field potential and the external potential, $V$is the volume, $q_1=2q>0$ is the charge of the ${\rm Ca}^{+2}$ ion, while $q_2=-q<0$ is the charge of ${\rm Cl}^{-}$. We have $N_2=2N_1$: the number of ${\rm Cl}^{-}$ ions is two times larger than the number of ${\rm Ca}^{+2}$
ions, as needed for electroneutrality. Note structural analogies between 
(\ref{poi1}) and (\ref{w7}). Eq.~(\ref{poi1}) can be derived using a similar mean-field method to (\ref{w7}). 

We assume that $\phi_{\rm ext}(\x)=\phi_{\rm ext}(r)$ in (\ref{poi1}) is spherically-symmetric ($r=|\x|$), and we look for a spherically symmetric $\phi(\x)=\phi(r)$. 
With notations $\varphi=\beta q \phi$, $\varphi_{\rm ext}=\beta q \phi_{\rm ext}$, $\varphi=\chi/r$, $\varphi_{\rm ext}=\chi_{\rm ext}/r$, $a=\sqrt{24\pi \beta q^2N_1/V}$, we get from (\ref{poi1}):
\be
\label{poi2}
&(\chi-\chi_{\rm ext})''+\frac{a^2 r}{3}(e^{-\chi/r}-e^{2\chi/r})=0,\\
&\chi(0)=\chi(\infty)=0,
\label{poi22}
\ee
where we note that $a$ is the inverse Debye length. Now $\chi(0)=0$ comes from the definition $\varphi(r)=\chi/r$, and from the fact that both $\varphi(0)$ and $\varphi_{\rm ext}(0)$ are finite.  
Condition $\chi(\infty)=0$ in (\ref{poi22}) comes from linearizing (\ref{poi2}) for a large $r$. This linearization produces Debye screening and leads to $\chi(\infty)=0$.

Note that in the nonlinear regime, (\ref{poi1}, \ref{poi2}) are not invariant with respect to the transformation (\ref{symsym}). However, what does this imply physically? Are the more numerous but less charged ${\rm Cl}^{-}$ ions better screeners for a positive external charge than less numerous but more charged ${\rm Ca}^{+2}$ ions?
The same question was asked in the previous section about charges that are not free, but are connected in a linear molecule; see Fig.~\ref{fig2}. As it turns out, the answers of 
Fig.~\ref{fig2} and Fig.~\ref{fig3} are opposite each other: Fig.~\ref{fig3} reports numerical solutions of (\ref{poi2}, \ref{poi22}) showing that more numerous and less charged mean-field interacting particles (i.e., ${\rm Cl}^-$ ions for the present case)
are better screeners. Appendix \ref{apa} provides a perturbative argument that validates these results. For a sufficiently small $c$ in $\chi_{\rm ext}(r)=c(1-e^{-br})$ [cf.~(\ref{ugol})], one can treat (\ref{poi2}) perturbatively:
\be
\label{akan}
\chi(r;c)\simeq \chi_0(r;c)+\chi_1(r;c^2),
\ee
where $\chi_0(r;c)$ solves exactly the linearized version of (\ref{poi2}), while 
$\chi_1(r;c^2)$ is the first-order correction. We have $\chi_0(r;c)=-\chi_0(r;-c)$ (as expected, since the ${\rm C}_{\rm eff}$-symmetry should hold for the linearized situation),
while $\chi_1(r;c^2)<0$; see Appendix \ref{apa}. Hence, 
\be
\label{aka}
\chi_++\chi_-=\chi(r;c)+\chi(r;-c)\simeq 2\chi_1(r;c^2)<0,
\ee
which is consistent with Fig.~\ref{fig3}. This figure shows that the message of (\ref{aka}) is valid as well for larger $c$, i.e., in the nonperturbative domain. For all values of $c$ we were able to check, $\varphi_+=\chi_+(r)/r$ is a monotonically decaying function of $r$; i.e., no effects similar to Fig.~\ref{fig01} are seen. 
We stress that the violation of the ${\rm C}_{\rm eff}$-symmetry for the nonlinear Poisson-Boltzmann do not require nonlocality. In other words, (\ref{poi1}) is a differential (not an integro-differential) equation; cf.~with our discussion around (\ref{yula}).

Interestingly, if the question before (\ref{akan}) is asked about the planar nonlinear Poisson-Boltzmann system -- i.e, now the negative and positive charged planes are screened -- then the answer is opposite to Fig.~\ref{fig3}; see Appendix \ref{apb}. Now ${\rm Ca}^{+2}$ ions are better screeners, which is similar to the result reported in Fig.~\ref{fig2}. 

Thus, there are different scenarios of ${\rm C}_{\rm eff}$-symmetry violation even for Poisson-Boltzmann systems.

\section{Summary}
\label{summary}

We studied mean-field dipolar systems, where molecular dipoles are extended, rigid linear structures with point-like charges, and the nonlinear situation applies in the vicinity of external charges. 
We explored new effects caused by a joint influence of nonlocality (due to dipole structure and finite size) and nonlinearity. 

Our first main result is that for dimer molecules -- the simplest situation, where the nonpoint-like structure is manifest -- the mean-field potential is nonmonotonic, indicating scenarios where like charges can attract each other. This effect arises from the solvation structure formed by molecules around a sufficiently strong external charge. 

Our second main result relates to the effective C-symmetry (${\rm C}_{\rm eff}$-symmetry) that is present in macroscopic electrodynamics. For nonpoint-like molecules this symmetry is violated due to nonlocal and (simultaneously) nonlinear effects in sufficiently complex liquids containing asymmetric molecules (the effect is absent for dimers due to their symmetry). In particular, the violation of the symmetry means that positive and negative (external) charges immersed into the electroneutral system of dipoles will be screened differently. We identified two mechanisms of this difference in screening and related it with the corresponding free energy differences. As expected, better screening relates to lower free energy. Our results show that it also relates with a higher entropy. To put these results in a broader perspective, we also studied violations of the ${\rm C}_{\rm eff}$-symmetry that are present in nonlinear Poisson-Boltzmann systems, which describe asymmetric electrolytes. These mechanisms allow (partially) analytical treatment due to relative simplicity of the Poisson-Boltzmann equation. We show that in such systems the asymmetric screening mechanism depends on the dimensionality, i.e., on whether a charged surface is screened or a spherical symmetric localized charge.

\section{Open problems and limitations of the present work}

This research raises several specific questions. They need to be addressed in future work. 

-- A major limitation of the present work is that the assumed rigid charge distribution of molecules neglects the induced charges (dipole moments), which are expected to be relevant, e.g., in the nonlinear situation. We are currently working to incorporate the effects of induced charges into the mean-field theory.

-- We need to understand more about the conditions under which the nonlinear equation (\ref{w91}) has well-defined solutions. One of the major differences between (\ref{w91}) and other mean-field nonlinear equations (e.g., the Poisson-Boltzmann equation (\ref{poi2})) is that (\ref{w91}) 
need not have guaranteed solutions, since it involves both advanced and retarded arguments \cite{driver}.

-- The attraction of like charges was shown in section \ref{attra} under a special condition: one of the external charges is rather strong and localized (such that it has a nontrivial solvation structure composed of dimer molecules), while another external charge is sufficiently weak so that it does not create its own solvation structure. We need to generalize this and provide a full theory of interacting charges with comparable magnitudes.

-- The issues of ${\rm C}_{\rm eff}$-symmetry violations and screening of positive versus negative external charges are intuitively appealing, and they perhaps should be addressed directly, beyond the mean-field limits. This problem relates to the well-known problem of finding the ground state of interacting dipoles on various lattices; see \cite{dipoles_thomson} for a recent review. It also relates to Thomson's problem of finding the ground state of interacting electrons; see e.g. \cite{thomson_recent,thomson_epl,podgornik} for recent discussions of this rich problem. 

-- The asymmetric screening studied in this work essentially depends on the characteristic length of the external charge (ion). This opens an interesting possibility of studying the rich physics of Hofmeister series \cite{hofmeister_good} (which includes the ion radius dependence of solvation) via nonlocal, nonlinear electrodynamics. A systematic study of this problem should involve accurate numerical calculations of free energy and entropy for the realistic range of parameters. 

-- The mean-field thermodynamics allows us to deduce a general expression not only for the mean energy, but also for the free energy; see section \ref{free}. There is therefore the possibility of constructing a more precise and interesting analogue of Born's theory of solvation; see \cite{born_model} for a recent critical account of this theory. 

-- In the present work, we restricted ourselves to linear molecules that cannot form hydrogen bonds (aprotic liquids). Even then, we did not explore the full range of realistic parameters. There are two reasons for this: the first reason has to do with the technical difficulties in solving integro-differential equations that contain both advanced and retarded arguments. We are currently working to design better numerical methods for such equations. Second, the proper mean-field method anyhow does not apply to protic liquids such as water. These arguments make clear why in the present paper we did not attempt the nonlocal, nonlinear electrostatic modeling of water; several problems are to be solved before such a modeling can be attempted. 

\section*{Acknowledgements}

The HESC of Armenia supported us under grants 24IRF/2-1C001 and 21AG-1C038.
We thank V.E. Mkrtchian, A. Hayrapetyan and D. Petrosyan for discussions.

\bibliography{nonlinear_references}

\appendix

\section{Validity of the mean-field approach and the magnitude of the dimensionless coupling constant $d$}
\label{validity}

The validity range of mean-field methods needs special attention
\cite{bavaud,kaufman}. One approach to the validity range starts with
the Curie-Weiss-Kac limit, where the thermodynamic limit for an
$N$-particle system is implemented via taking $N\to\infty$
simultaneously with weakening the two-body interaction potential as
$\propto \frac{1}{N}$ \cite{bavaud}; i.e.,
$V(Q_i,Q_j)\to\frac{1}{N}V(Q_i,Q_j) $ in (\ref{mok1}). In this limit, the
mean-field approach is exact \cite{bavaud}. This is useful for producing
exactly-solvable models, but it does not explain the applicability of
the mean-field method to a real system, where two-body interactions
do not scale as $\propto \frac{1}{N}$. Another way to approach
the validity range is to employ exact and approximate formulas of
equilibrium statistical mechanics, where the one-particle probability
$P(x)$ is formally given by (\ref{6}), but the effective energy $U_{\rm
eff}(x)$ in (\ref{6}) depends not only on $V_{\rm ext}(x)$ and $V(x,y)$
[as in (\ref{7})], but also on the two-particle correlation function
\cite{bavaud,kaufman}. 

Here we follow a more intuitive method for defining the validity range
of the mean-field approach \cite{debye}. Ref.~\cite{kaufman}
shows that similar results are found from the above correlation
function method. Mean-field methods can be valid when collisions
between two particles are irrelevant in comparison to many-body effects
\cite{debye,kaufman}. To explain the meaning of parameter $d$ in
(\ref{ono}), and get some ideas on the applicability of the mean field,
we shall apply the general qualitative argument of \cite{debye} to
dipolar gases. 

A strong, two-particle collision means that the potential energy
$\frac{q^2a^2}{\varepsilon r^3_0}$ for two interacting dipoles (dimers)
reaches their relative kinetic energy, which in equilibrium is
estimated via $\sim T$. Hence, we can define such a collision as
$\frac{q^2a^2}{\varepsilon r^3_0}\sim T$, where $r_0$ is the collision
distance \cite{debye}. Such collisions are rare (and hence the
mean-field applies to thermodynamic quantities), if $r_0^3\ll V/N$,
where $(V/N)^{1/3}$ is the mean inter-particle distance. Now $r_0^3\ll
V/N$ can be satisfied in two different cases. The first case is
$\frac{N\beta q^2|\a|^2}{V}\ll 1$; cf.~(\ref{ono}). This is the
non-interesting limit of almost non-interacting dipoles. The second case
is when $\varepsilon=1+\frac{16d}{3}\simeq \frac{16d}{3}$; cf.~(\ref{khan}). Then $r_0^3\ll V/N$ reduces to $3/16\ll 1$. This is a stretched condition, but it indicates the
qualitative applicability of the mean-field method to moderately coupled
dipoles. 

To conclude, we would like to clarify why we chose long-range couplings $\frac{q^2a^2}{\varepsilon r^3_0}$ between two dipoles. The reason is that when $r_0^3$ is estimated via $\frac{q^2a^2}{\varepsilon r^3_0}\sim T$, we get $r_0^3\gg a^3$ for typical molecular sizes. The meaning of this is that the mean-field is a collective effect, and its validity is determined from long-range effects.

\section{Free energy, mean energy and entropy for mean-field electrostatics}
\label{free_ap}

We start by outlining the mean-field thermodynamics using the generalized coordinate $Q$; 
cf.~(\ref{mok1}, \ref{gibarian}). Later, it will be applied to Coulomb interactions within nonlocal, nonlinear electrostatics. We leave aside the nonequilibrium thermodynamics of mean-field systems. This interesting subject goes beyond the scope of the present work. 

\subsection{Thermodynamics of mean-field systems}

Note from (\ref{mok1}, \ref{gibarian}) that the mean-field approximation for the stationary average energy is to be defined as 
\begin{align}
&\overline{U}=\int\prod_{I=1}^N\d Q_I\, P(Q_I)\,U(Q_1,...,Q_N)\nonumber\\
&=N\int\d Q\, P(Q) V_{\rm ext}(Q)\\
&+\frac{N(N-1)}{2}
\int\d Q_1\,\d Q_2 P(Q_1)P(Q_2)
V(Q_1,Q_2).
\label{burundi}
\end{align}
Using (\ref{6}, \ref{7}), we express $\overline{U}$ via $U_{\rm eff}$ and $V_{\rm ext}$:
\begin{align}
    \overline{U}= \frac{N}{2}\int\d Q\, P(Q) [V_{\rm ext}(Q)+U_{\rm eff}(Q)].
\label{oko1}
\end{align}
The free energy within the mean-field approach is defined as 
\begin{align}
\label{oko2}
&F= \overline{U}-TN\Big[-\int\d x\, P(Q)\ln P(Q)   \Big]
\\
&= -NT\ln Z
+\frac{N}{2}\int\d Q\, P(Q) [V_{\rm ext}(Q)-U_{\rm eff}(Q)],
\label{oko3}
\end{align}
where we used (\ref{oko1}) when deriving (\ref{oko3}) from (\ref{oko2}). 
The rationale of (\ref{oko2}) is that the real (non-mean-field) free energy 
${\cal F}$ (deduced from (\ref{gibarian})) is bounded by $F$ (Bogolyubov's inequality) [cf.~(\ref{mok1})]:
\begin{align}
\label{oko4}
&{\cal F}\leq F,\\
&{\cal F}=-T\ln {\cal Z}, \quad 
{\cal Z}=\int\prod_{I=1}^N\d Q_I\,e^{-\beta U(Q_1,...,Q_N) }.
\end{align}
Eq.~(\ref{oko4}) follows from the positivity of the relative entropy
\begin{align}
\label{oko5}
&-\int\prod_{I=1}^N\d Q_I\, \hat {\cal P}\ln  \hat {\cal P}
\leq -\int\prod_{I=1}^N\d x_k\, \hat {\cal P}\ln  {\cal P}, \\
&\hat {\cal P}(Q_1,...,Q_N)=\prod_{k=1}^N P(Q_k),\\
&{\cal P}(Q_1,...,Q_N)=\frac{1}{{\cal Z}} e^{-\beta U(Q_1,...,Q_N) }.
\end{align}

So far $F$ was introduced only via the Bogolyubov inequality. We now show that it describes the mean-field approximation
for the isothermal work. This fully legitimizes the meaning of $F$.

We assume that there is a slowly varying parameter $\lambda$, which enters (\ref{mok1}). 
Differentiate (\ref{oko3}) over $\lambda$ and employ (\ref{7}):
\begin{align}
\label{ra1}
2\partial_\lambda F-N\langle \partial_\lambda U_{\rm eff}\rangle=N\langle \partial_\lambda V_{\rm ext}\rangle
-\frac{N(N-1)}{2}\int\d Q_1\d Q_2\, V(Q_1,Q_2)\partial_\lambda [P(Q_1)P(Q_2)],
\end{align}
where $\langle ...\rangle$ means averaging over the stationary $P(Q)$. Now differentiating (\ref{7}) over $\lambda$ and averaging it we find
\begin{align}
\langle \partial_\lambda U_{\rm eff}\rangle&=\langle \partial_\lambda V_{\rm ext}\rangle+(N-1)
\int\d Q_1\d Q_2 \partial_\lambda [V(Q_1,Q_2)] P(Q_1)P(Q_2)\nonumber\\
&+\frac{N-1}{2}\int\d Q_1\d Q_2\, V(Q_1,Q_2)\partial_\lambda [P(Q_1)P(Q_2)].
\label{ra2}
\end{align}
Now multiplying (\ref{ra2}) by $N$, adding it to (\ref{ra1}),
and noting that the slow isothermal work is defined as
\begin{align}
\label{ra3}
\frac{\delta W}{\delta \lambda}
=N\langle \partial_\lambda V_{\rm ext}\rangle+\frac{N(N-1)}{2}
\int\d Q_1\d Q_2 \partial_\lambda [V(Q_1,Q_2)] P(Q_1)P(Q_2),
\end{align}
we get (as we wanted to show)
\begin{align}
\label{ra4}
\partial_\lambda F=\frac{\delta W}{\delta \lambda}.
\end{align}

Multiplying (\ref{ra2}) by $-N$, and adding it to (\ref{ra1}), we get another useful relation:
\begin{align}
\label{ra44}
&\partial_\lambda F-N\langle \partial_\lambda U_{\rm eff}\rangle=
-\partial_\lambda\langle U_{\rm int}\rangle, \\
&\langle U_{\rm int}\rangle = 
\frac{N(N-1)}{2}\int\d Q_1\d Q_2\, V(Q_1,Q_2)P(Q_1)P(Q_2),
\end{align}
where $\langle U_{\rm int}\rangle$ is the mean interaction energy; see also (\ref{burundi}).

\subsection{Application to electrostatics}
\label{ella}

We recall (\ref{w1}--\ref{w7}) and start with (\ref{oko1}) and (\ref{v5}, \ref{w5}):
\be
\bar{U}=\frac{N}{2Z}\int_V \d \y\oint\d \b\, e^{-\beta \T(\y,\b)\phi(\y)} \T(\y,\b)[\phi(\y)+\phi_{\rm ext}(\y)].
\label{w8}
\ee
The integral $\int_V \d \y$ in (\ref{w8}) is taken by parts; thereby $\T(\y,\b)$ turns into $\T(\y,-\b)$. The boundary terms during the integration by parts are neglected due to the thermodynamic limit. We apply (\ref{w7}) and integrate by parts again (this last integration by parts does not have a boundary term, since the integral can be extended to infinity):
\be
\bar{U}=\frac{1}{8\pi}\int \d \y \, [\na\phi(\y)-\na\phi_{\rm ext}(\y)]
[\na\phi(\y)+\na\phi_{\rm ext}(\y)]=\frac{1}{8\pi}\int \d \y \, [(\na\phi(\y))^2-(\na\phi_{\rm ext}(\y))^2].
\label{w9}
\ee
The meaning of (\ref{w9}) is clear: this is the full electric energy minus the energy of the external field. Note that (\ref{w9}) applies to the nonlinear situation. Likewise, we have for the free energy from (\ref{oko3}):
\be
F=-TN\ln Z-\frac{1}{8\pi}\int \d \y \, [\na\phi(\y)-\na\phi_{\rm ext}(\y)]^2.
\label{w10}
\ee
Macroscopic electrodynamics does not have a closed expression for the free energy $F$; see (\ref{w10}, \ref{oko3}). Instead, it offers an expression for $\partial_\lambda F_{\rm macro}$ \cite{landau8}, where $\lambda$ is a parameter of the system. One can show that 
$\partial_\lambda F_{\rm macro}=\partial_\lambda F$, where $F$ is given by (\ref{w10}).

To clarify the meaning of (\ref{w10}), note from (\ref{w3}) that the dipole moment of a single molecule is
\be
\sum_{j=1}^nq_j\a_j=\T(\y,\a)\y.
\ee
Hence the full average dipole moment is
\be
\bar{{\bm{P} }}=\frac{N}{Z}\int_V \d \y\oint\d \b\, e^{-\beta \T(\y,\b)\phi(\y)} \T(\y,\b)\y.
\label{w11}
\ee
Integrating by parts (boundary terms are neglected due to the thermodynamic limit), and employing (\ref{w7}), we find
\be
\bar{{\bm{P} }}=\frac{1}{4\pi}\int_V \d \y \, [\na\phi(\y)-\na\phi_{\rm ext}(\y)],
\ee
which is similar to the usual formula of macroscopic electrodynamics \cite{landau8}.

\section{Perturbative solution of Eq.~(\ref{poi2})}
\label{apa}

We expand (\ref{poi2}) assuming that $\chi$ is small and retaining only the first nonlinear term:
\be
\label{poi3}
&(\chi-\chi_{\rm ext})''-a^2 \chi -\frac{a^2}{2r} \chi^2=0,\quad r\geq 0, \\
&\chi(0)=\chi(\infty)=0,
\label{poi4}
\ee
where (\ref{poi4}) are the standard boundary conditions. 
To solve (\ref{poi3}) we first neglect the nonlinear term and solve:
\be
\label{poi3}
(\chi_0(r)-\chi_{\rm ext}(r))''-a^2 \chi_0(r) =0,
\ee
via the Green's function method under boundary conditions 
$\chi_0(0)=\chi_0(\infty)=0$:
\be
\chi_0(r)=-\frac{1}{2a}\int_0^\infty \d s\, [e^{-a|s-r|}-e^{-a(s+r)}]\,\chi_{\rm ext}''(s). 
\label{ams1}
\ee
In (\ref{ams1}) we employ 
$\chi_{\rm ext}''(r) =-c b^2 e^{-br}$ given by (\ref{ugol}). Due to $e^{-a|s-r|}-e^{-a(s+r)}\geq 0$ we obviously find $\chi_0(r)\geq 0$, as confirmed by an explicit formula found from (\ref{ams1}):
\be
\chi_0(r)=\frac{cb^2}{b^2-a^2}[e^{-ar}-e^{-br}].
\ee
For the first-order nonlinear correction $\chi_1(r)$ we get from (\ref{poi3}, \ref{ams1}):
\be
\label{poi33}
&&\chi_0(r)''-a^2 \chi_0(r) =\frac{a^2}{2r} \chi_0^2,\\
&&\chi_1(r)=-\frac{1}{2a}\Big[\frac{cb^2}{b^2-a^2} \Big]^2
\int_0^\infty \frac{\d s}{s}\, [e^{-a|s-r|}-e^{-a(s+r)}]\,
[e^{-as}-e^{-bs}]^2.
\label{ams2}
\ee
Even without looking at the explicit expression (\ref{ams2}), we note that $\chi_1(r)<0$ due to $e^{-a|s-r|}-e^{-a(s+r)}\geq 0$ and that $\chi_1(r)$ is an even function of $c$. 

\section{Solution of planar Poisson-Boltzmann equation}
\label{apb}

Here we study the 1d (planar) version of (\ref{poi1}) [$\varphi''(x)\equiv\frac{\d^2}{\d x^2}\varphi$]:
\be
\label{poi11}
&&\varphi''(x)+\hat a^2\sum_{\alpha=1}^{{\cal N}}\hat q_\alpha \hat n_\alpha e^{-\hat q_\alpha\varphi}=0,\\
&& x>0,\qquad \sum_{\alpha=1}^{{\cal N}}\hat q_\alpha \hat n_\alpha=0.
\label{mrmr}
\ee
Eq.~(\ref{poi11}) describes ${\cal N}$ different types of ions in volume $V$; their numbers and charges are (resp.) $N_\alpha=\hat n_\alpha N$ and
$\hat q_\alpha q$, where $\hat n_\alpha$ are positive integers, $\hat q_\alpha$ are positive or negative integers, $\hat a=\sqrt{4\pi \beta q^2N/V}$ is the inverse Debye length, and $\varphi=\beta q\phi$ is the dimensionless potential. $\sum_{\alpha=1}^{{\cal N}}\hat q_\alpha \hat n_\alpha=0$ in (\ref{mrmr}) refers to electroneutrality. 

The external field is introduced via boundary conditions: we assume a conducting plane located at $x=0$. The fixed surface charge density $\sigma$ determines 
\be
\label{bou}
\varphi'(0)= -4\pi \beta q\sigma.
\ee
The other boundary condition $\varphi(\infty)=0$ comes from linearizing (\ref{poi11}) for larger distances and using (\ref{mrmr}). Now (\ref{poi11}) is formally a conservative Newton equation for the ``coordinate'' $\varphi$ and ``time'' $x$. Its energy conservation together with $\varphi(\infty)=0$ leads to
\be
\label{poi33}
[\varphi'(x)]^2=\psi(x), \quad x\geq 0, \\
\psi(x)\equiv 2\hat a^2\sum_{\alpha=1}^{{\cal N}} \hat n_\alpha 
[e^{-\hat q_\alpha\varphi(x)}-1], 
\ee
We get from (\ref{bou}, \ref{poi33}):
\be
\label{bars1}
\varphi'(x)=-\sqrt{\psi(x)}, \quad {\rm for}\quad \varphi(0)>0\quad {\rm and}\quad \varphi'(0)<0,\\
\varphi'(x)=\sqrt{\psi(x)}, \quad {\rm for}\quad \varphi(0)<0\quad {\rm and}\quad \varphi'(0)>0.
\label{bars2}
\ee
Now consider ${\rm Ca}^{+2}+2{\rm Cl}^{-}$, i.e., put in (\ref{poi11}):
\be
{\cal N}=2,\quad \hat n_1=1,\quad \hat q_1=2,\quad \hat n_2=2,\quad \hat q_2=-1.
\ee
Denote solutions of (\ref{bars1}) and (\ref{bars2}) as (resp.) $\varphi_+$ and $\varphi_-$, and take 
\be
\varphi_+'(0)<0,\quad
\varphi_-'(0)>0,\quad
\varphi_+'(0)=-\varphi_-'(0),
\ee
i.e., the two solutions $\varphi_+$ and $\varphi_-$ refer to the charged surfaces having (resp.) positive and negative surface charge $\pm\sigma$ in (\ref{bou}). Hence $\varphi_+(0)$ and $\varphi_-(0)$ are two different (positive and negative, respectively) solutions of the same equation (\ref{poi33}):
\be
\label{brbr}
[\varphi_{\pm}'(0)]^2=2\hat{a}^2[e^{-2\varphi_{\pm}(0)}+2e^{\varphi_{\pm}(0)}-3].
\ee
The inequality 
\be
2e^{-y}+e^{2y}>2e^{y}+e^{-2y}, \quad y>0, 
\label{bor}
\ee
implies: 
$\varphi_+(0)>-\varphi_-(0)$ from (\ref{brbr}). Now two monotonous, positive functions $\varphi_+(x)>-\varphi_-(x)$ hold
$\varphi_+(0)>-\varphi_-(0)$ and $\varphi'_+(0)=-\varphi'_-(0)$. The numerical solutions confirm that
\be
\label{bro}
\varphi_+(x)>-\varphi_-(x), \quad x\geq 0,
\ee
which is traced back to inequality (\ref{bor}).

Eq.~(\ref{bro}) means that the positive plane is screened less than the negative surface; i.e., fewer ${\rm Ca}^{+2}$ ions are better screeners than more numerous (but less charged) ${\rm Cl}^{-}$ ions.

\end{document}